\pdfoutput=1
\documentclass[11pt,twoside,a4paper,cmspaper,final,collab]{cms-tdr}
\def\svnVersion{cc7bf21}\def\svnDate{2026/01/26}\def\cmsCernNoTag{CERN-EP-2025-162}\def\cmsCernDate{\today}\def\cmsMessage{Published in Physical Review Letters as \href{http://dx.doi.org/10.1103/njf9-4zfv}{\doi{10.1103/njf9-4zfv}.}}
\begin{document}\cmsNoteHeader{BPH-24-011}

\newcommand{\mumu}      {\ensuremath{\PGmp\PGmm}\xspace}
\newcommand{\piz}       {\ensuremath{\PGpz}\xspace}
\newcommand{\Kstarzm}   {\ensuremath{\HepParticleResonanceFull{\PK}{}{*}{892}{}{0}}\xspace}
\newcommand{\Kstarz}    {\ensuremath{\HepParticle{\PK}{}{\ast0}}\Xspace}
\newcommand{\Bq}        {\ensuremath{\HepParticle{\PB}{}{+,0}}\xspace}
\newcommand{\Bqst}      {\ensuremath{\HepParticle{\PB}{}{*+,0}}\xspace}
\newcommand{\mBu}       {\ensuremath{m(\PBp)}\xspace}
\newcommand{\mBd}       {\ensuremath{m(\PBz)}\xspace}
\newcommand{\mBs}       {\ensuremath{m(\PBzs)}\xspace}
\newcommand{\Bst}       {\ensuremath{\PBst}\xspace}
\newcommand{\Bust}      {\ensuremath{\PBstp}\xspace}
\newcommand{\Bdst}      {\ensuremath{\PBstz}\xspace}
\newcommand{\Bsst}      {\ensuremath{\PBstzs}\xspace}
\newcommand{\mB}        {\ensuremath{m(\PB)}\xspace}
\newcommand{\mBst}      {\ensuremath{m(\PBst)}\xspace}
\newcommand{\mBust}     {\ensuremath{m(\PBstp)}\xspace}
\newcommand{\mBdst}     {\ensuremath{m(\PBstz)}\xspace}
\newcommand{\mBsst}     {\ensuremath{m(\PBstzs)}\xspace}
\newcommand{\dmBst}     {\ensuremath{\Delta m(\PBst)}\xspace}
\newcommand{\dmBust}    {\ensuremath{\Delta m(\PBstp)}\xspace}
\newcommand{\dmBdst}    {\ensuremath{\Delta m(\PBstz)}\xspace}
\newcommand{\dmBsst}    {\ensuremath{\Delta m(\PBstzs)}\xspace}
\newcommand{\Bg}        {\ensuremath{\PB\PGg}\xspace}
\newcommand{\Bug}       {\ensuremath{\PBp\PGg}\xspace}
\newcommand{\Bdg}       {\ensuremath{\PBz\PGg}\xspace}
\newcommand{\Bsg}       {\ensuremath{\PBzs\PGg}\xspace}
\newcommand{\BstBg}     {\ensuremath{\Bst\to\Bg}\xspace}
\newcommand{\BsstBsg}   {\ensuremath{\Bsst\to\Bsg}\xspace}
\newcommand{\jpsi}      {\ensuremath{\PJGy}\xspace}
\newcommand{\psit}      {\ensuremath{\PGyP{\mathrm{2S}}}\xspace}
\newcommand{\KK}        {\ensuremath{\PKp\PKm}\xspace}
\newcommand{\PKK}       {\ensuremath{\PGf\to\KK}\xspace}
\newcommand{\mbpdg}     {\ensuremath{M^{\text{PDG}}_{\PB}}\xspace}
\newcommand{\pp}        {\ensuremath{\Pp\Pp}\xspace}
\newcommand{\PsiK}      {\ensuremath{\PGy\PKp}\xspace}
\newcommand{\PsiKst}    {\ensuremath{\PGy\Kstarzm}\xspace}
\newcommand{\PsiKstnom} {\ensuremath{\PGy\Kstarz}\xspace}
\newcommand{\PsiKpi}    {\ensuremath{\PGy\PKp\PGpm}\xspace}
\newcommand{\PsiKK}     {\ensuremath{\PGy\PKp\PKm}\xspace}
\newcommand{\PsiPhi}    {\ensuremath{\PGy\PGf}\xspace}
\newcommand{\BuJpsiPi}  {\ensuremath{\PBp\to\jpsi\PGpp}\xspace}
\newcommand{\BuPsiK}    {\ensuremath{\PBp\to\PsiK}\xspace}
\newcommand{\BuPsiPi}   {\ensuremath{\PBp\to\PGy\PGpp}\xspace}
\newcommand{\BdPsiKst}  {\ensuremath{\PBz\to\PsiKst}\xspace}
\newcommand{\BdPsiKstnom}{\ensuremath{\PBz\to\PsiKstnom}\xspace}
\newcommand{\BdPsiKPi}  {\ensuremath{\PBz\to\PsiKpi}\xspace}
\newcommand{\BsPsiPhi}  {\ensuremath{\PBzs\to\PsiPhi}\xspace}
\newcommand{\ee}        {\ensuremath{\Pep\Pem}\xspace}
\newcommand{\GG}        {\ensuremath{\PGg\PGg}\xspace}
\newcommand{\PizGG}     {\ensuremath{\piz\to\GG}\xspace}
\newcommand{\chisq}     {\ensuremath{\chi^2}\xspace}
\newcommand{\pvtx}      {\ensuremath{P_{\text{vtx}}}\xspace}
\newcommand{\ptet}      {\ensuremath{\pt\text{ and }\abs{\eta}}\xspace}
\newcommand{\ptg}       {\ensuremath{\pt(\PGg)}\xspace}
\newcommand{\aeta}      {\ensuremath{\abs{\eta}}\xspace}
\newcommand{\aetag}     {\ensuremath{\abs{\eta(\PGg)}}\xspace}
\newcommand{\mBq}       {\ensuremath{m(\Bq)}\xspace}
\newcommand{\mBqst}     {\ensuremath{m(\Bqst)}\xspace}
\newcommand{\dmBqst}    {\ensuremath{\Delta m(\Bqst)}\xspace}

\newlength\cmsTabSkip\setlength{\cmsTabSkip}{1ex}
\providecommand{\cmsTable}[1]{\resizebox{\textwidth}{!}{#1}}
\ifthenelse{\boolean{cms@external}}
{\providecommand{\cmsTabla}[1]{\resizebox{0.93\textwidth}{!}{#1}}}
{\providecommand{\cmsTabla}[1]{\resizebox{\textwidth}{!}{#1}}}

\ifthenelse{\boolean{cms@external}}
{\providecommand{\suppMaterial}{The Supplemental Material~\cite{suppMatBib}}}
{\providecommand{\suppMaterial}{Appendix~\ref{app:suppMat}\nocite{suppMatBib}}}

\ifthenelse{\boolean{cms@external}}
{\providecommand{\suppMaterialB}{the Supplemental Material~\cite{suppMatBib}}}
{\providecommand{\suppMaterialB}{Appendix~\ref{app:suppMatB}\nocite{suppMatBib}}}

\ifthenelse{\boolean{cms@external}}
{\providecommand{\suppMaterialC}{the Supplemental Material~\cite{suppMatBib}}}
{\providecommand{\suppMaterialC}{Appendix~\ref{app:suppMatC}\nocite{suppMatBib}}}

\cmsNoteHeader{BPH-24-011}
\title{First exclusive reconstruction of the \texorpdfstring{$\Bust$}{B*+}, \texorpdfstring{$\Bdst$}{B*0}, and \texorpdfstring{$\Bsst$}{Bs*0} mesons and precise measurement of their masses}

\date{\today}

\abstract{
Using proton-proton collision data collected by the CMS experiment at $\sqrt{s}= 13\TeV$ in 2016--2018, corresponding to an integrated luminosity of $140\fbinv$, the first full reconstruction of the three vector \PB meson states, \Bust, \Bdst, and \Bsst, is performed. The mass differences between the excited mesons and their corresponding ground states are measured to be $\mBust-\mBu = 45.277\pm0.039\pm0.027\MeV$, $\mBdst-\mBd = 45.471\pm0.056\pm0.028\MeV$, and $\mBsst-\mBs = 49.407\pm0.132\pm0.041\MeV$, where the first uncertainties are statistical and the second are systematic. These results improve on the precision of previous measurements by an order of magnitude.
}
\hypersetup{%
pdfauthor={CMS Collaboration},%
pdftitle={First exclusive reconstruction of the B*+, B*0, and Bs*0 mesons and precise measurement of their masses},%
pdfsubject={CMS},%
pdfkeywords={CMS, heavy flavor, flavor spectroscopy, experimental results, radiative transitions}}

\maketitle 

While the ground states of beauty mesons \PBp, \PBz, and \PBzs have been extensively studied, 
the corresponding excited vector mesons \Bust, \Bdst, and \Bsst are much less explored. The main challenge 
in the reconstruction of these excited states stems from the low-energy (40--50\MeV in the \Bst meson rest frame)
photons emitted in \BstBg decays, with the energy below the detection thresholds in most experiments.
Throughout this Letter, \Bst corresponds to one of the \Bust, \Bdst, or \Bsst mesons, and
\PB refers to one of the \PBp, \PBz, or \PBzs mesons; the inclusion of charge-conjugate states is implied.
About 30 years ago, the four LEP experiments 
combined \PQb jets with photons reconstructed from their conversions into \ee pairs to measure the flavor-averaged 
mass difference $\mBst-\mB$~\cite{L3:1994jps,ALEPH:1995ikc,DELPHI:1995hfn,OPAL:1996ecm}, also known as the hyperfine splitting. 
Studies of the P-wave \PBzs mesons decaying into $\PBp\PKm$ and $\Bust\PKm$ by the LHCb and CMS 
experiments~\cite{LHCb:2012iuq, CMS:2018wcx}, as well as to $\PBz\PKzS$ and $\Bdst\PKzS$ by CMS~\cite{CMS:2018wcx},
provided the measurements of the mass differences $m(\Bust)-m(\PBp)$ and $m(\Bdst)-m(\Bust)$ 
without reconstructing low-energy photons. 
Measurements at the dedicated \PB physics experiments CLEO and Belle, at the \PgUe energy,
allowed indirect estimates of the \Bsst meson mass~\cite{CLEO:2006yez,Belle:2008ezn}; 
however, these two results are in tension with each other. Measurements of the properties 
of excited \PB mesons provide an important input to our understanding of the quantum 
chromodynamics~(QCD) mechanisms responsible for quark dynamics and the formation of hadrons. 
In particular, precise measurements of the hyperfine splitting in the \PB meson system are an important 
input for quark models of hadrons~\cite{Rosner:1992qw,Goity:2007fu,Karliner:2019lau}.

In this Letter, the first exclusive reconstruction of the three excited vector
mesons \Bust, \Bdst, and \Bsst, is presented. The analysis is performed using 
proton-proton~($\pp$) collision data recorded by 
the CMS experiment during LHC Run 2 in 2016--2018 at $\sqrt{s}=13\TeV$, corresponding to an 
integrated luminosity of 140\fbinv~\cite{Sirunyan:2759951,CMS-PAS-LUM-17-004,CMS-PAS-LUM-18-002}. 
Low-energy photons are reconstructed through their conversions into \ee pairs in the beam pipe and detector material. 
Ground-state \PB mesons are reconstructed using the decays 
\BuPsiK, \BdPsiKstnom, and \BsPsiPhi, where \PGy stands for the \jpsi and \psit mesons detected 
using their decays to \mumu, and \Kstarz and \PGf refer to the \Kstarzm and $\PGf(1020)$ mesons, 
reconstructed through their decays to $\PKp\PGpm$ and \KK, respectively.
Tabulated results are provided in the HEPData record for this analysis~\cite{hepdata}.

The CMS apparatus~\cite{Chatrchyan:2008zzk,CMSRun3} is a multipurpose, nearly hermetic detector, 
designed to trigger on~\cite{Khachatryan:2016bia,CMS:2020cmk,CMS:2024aqx} and identify electrons, muons, 
photons, and charged and neutral hadrons~\cite{CMS:2020uim,CMS:2018rym,CMS:2014pgm}. 
A global ``particle-flow'' algorithm~\cite{CMS:2017yfk} aims to reconstruct all individual 
particles in an event, combining information provided by the all-silicon tracker and by the 
crystal electromagnetic and brass-scintillator hadron calorimeters, operating inside a 3.8\unit{T} 
superconducting solenoid, with data from the gas-ionization muon detectors embedded in the 
flux-return yoke outside the solenoid. Tracks are reconstructed from hits in the tracker and muons 
are reconstructed by matching tracks with muon candidates found in the muon detectors.
The procedure followed for aligning the detector is described in Ref.~\cite{CMS:2022ime}. 
An improved pixel detector was installed before the 2017 data taking began~\cite{Phase1Pixel}.

Events of interest are selected using a two-tiered trigger system~\cite{Khachatryan:2016bia}. 
The first level~(L1), composed of custom hardware processors, uses information from the 
calorimeters and muon detectors to select events at a rate of around 100\unit{kHz} within a fixed latency 
of 4\mus~\cite{CMS:2020cmk}. The second level, known as the high-level trigger 
(HLT), consists of a farm of computing processors running a version of the full event 
reconstruction software optimized for fast processing, and reduces the event rate to around 
1\unit{kHz} before data storage. The events used in this analysis were selected at L1 by requiring 
the presence of at least two muons, and at the HLT by requiring two opposite-sign~(OS) muons, 
with various pseudorapidity~$(\eta)$ and transverse momentum~(\pt) thresholds,
compatible with being produced in dimuon decays of either the \jpsi or \psit meson.

The offline reconstruction requires two OS muons that form a vertex
with fit \chisq probability $\pvtx>1\%$, each satisfying $\pt>3.5\gev$ and $\aeta<2.5$ 
requirements, and passing a special selection optimized for low-\pt muons~\cite{Chatrchyan:2012xi}. 
The dimuon invariant mass $m(\mumu)$ must be within 2.9--3.4\GeV for the \jpsi decay channels, 
and 3.4--3.9\GeV for the \psit decays.
To build \PBp (\PBz and \PBzs) candidates, the selected \PGy candidate is combined with one (two OS)
high-purity tracks~\cite{CMS:2014pgm} with $\pt>1\gev$ that do not match the selected muons.
For the \BuPsiK and \BsPsiPhi selection, one or both tracks are assigned the
charged kaon mass hypothesis, with an additional requirement of 
$1.01 < m(\KK) < 1.03\GeV$ for the latter.
For the \BdPsiKst selection, the $\PKp\PGpm$ or $\PKm\PGpp$ mass assignment is 
made based on which combination has an invariant mass closer to the world-average \Kstarz meson 
mass of $M^{\text{PDG}}_{\Kstarz}=895.55\MeV$~\cite{PDG}, followed by the requirement 
$0.81 < m(\PKpm\PGpmp) < 0.97\gev$.
In addition, a \PKK veto is applied in the \BdPsiKst channel (by changing the track mass hypothesis) 
to ensure that the \Bsg and \Bdg samples do not overlap.

The \PBp (\PBz and \PBzs) candidates are obtained via a three-track (four-track) vertex fit of the 
selected muons and track(s), which constrains the dimuon invariant mass to 
the world-average \jpsi or \psit mass~\cite{PDG}.
From all reconstructed \pp collision points, the primary vertex~(PV) is chosen as the one 
with the smallest \PB pointing angle, defined as the angle between 
the \PB candidate momentum and the vector connecting the PV and the reconstructed \PB candidate vertex.
Furthermore, if any of the three or four tracks used in the 
\PB candidate reconstruction is included in the fit of the chosen 
PV, they are removed and the PV position is refitted. 
The \PB candidates are required to have $\pt(\PB)>10\gev$, $\pvtx(\PB)>1\%$, 
transverse vertex displacement from PV $L_{xy}(\PB)$ larger than 3 times its uncertainty, 
two-dimensional pointing angle $\cos[\vec{L}_{xy}(\PB)$, $\vec{\pt}(\PB)]> 0.999$, 
and invariant mass within twice the average mass resolution of 
the corresponding \PB meson world-average mass \mbpdg~\cite{PDG}.

Photon conversion candidates are formed from two OS
tracks with a small angular separation, a small distance of closest approach,
a conversion vertex at least 1.5\cm away from the beam axis,
and a \chisq probability above 0.05\% for a kinematic fit that constrains the 
two tracks to originate from a common vertex with an invariant mass of zero.
More details on the selection criteria are given in Ref.~\cite{BPH13005}.
Only photon candidates with $\pt > 300\MeV$ and $\aeta < 1.5$ are kept, since 
conversions at more forward pseudorapidity are reconstructed with a significantly worse resolution.
To improve the resolution, the \Bg invariant mass is evaluated using the expression 
$M(\Bg) = m(\Bg)-m(\PB)+\mbpdg$, where the \PB meson and the 
conversion photon four-momenta are taken from the kinematic vertex fit that 
constrains the \PB meson, the photon candidate, and all tracks forming the chosen PV, 
to originate from the same vertex~\cite{BPH19003,BPH20004,BPH23002}. 

The momenta of the \Pep and \Pem tracks measured by the tracker tend to be lower than the 
true momenta because the electrons and positrons lose energy while traversing the tracker, 
which results in a measured photon energy lower on average than the true photon energy~\cite{BPH17008,BPH13005}. 
A photon energy scale~(PES) correction is performed 
using \PizGG decays reconstructed from two converted photons in several ranges of 
photon \ptet. 
The energy correction factor in each range is given by the ratio between the nominal \piz mass~\cite{PDG} 
and the mean \piz mass in data obtained from fitting the \GG invariant mass distribution. 
The PES correction is 1.002--1.04 in the kinematic range relevant for this analysis, \ie, $\ptg<1.5\gev$. 
\suppMaterial{} contains example fits to the \GG invariant mass distribution before and after the PES correction.

The \PYTHIA~8.240 package~\cite{Pythia8} with the CP5 underlying event tune~\cite{CMS:2019csb}
is used to simulate the \Bst meson production.
The \EVTGEN~1.6.0~\cite{Lange:2001uf} program models the decays \BstBg, \BuPsiK, 
\BdPsiKst, and \BsPsiPhi followed by the $\PGy\to\mumu$ decays~\cite{PDG}.
Final-state photon radiation is described using \PHOTOS~3.61~\cite{PHOTOS,PHOTOS_2}.
The generated Monte Carlo~(MC) events are then passed to a detailed \GEANTfour-based 
simulation~\cite{Agostinelli:2002hh} of the CMS detector. The simulated events are then processed 
by the same reconstruction algorithms as used for the collision data. The simulation includes the 
effects of multiple \pp interactions in the same or nearby bunch crossings~(pileup) with a multiplicity 
distribution matching that observed in data. The simulated \pt and \aeta distributions of all 
three \Bst mesons were adjusted using the corresponding \Bust meson distributions measured in data, 
as the data-simulation differences in the kinematic spectra are expected to be similar for the three states.

The \Bg mass resolution is found to depend strongly on \aetag, ranging from 
about 0.7\MeV at \aetag close to zero to about 2\MeV at \aetag around 1.5.
To improve the accuracy of the \Bst meson mass measurements, a simultaneous fit 
of the \Bg mass distributions in several categories, defined by ranges of \aetag, is performed. 
Pseudo-experiments were used to obtain a set of optimal \aetag ranges for each meson flavor. 
The resulting simultaneous fit structure includes:
three \aetag ranges with boundaries at 0, 0.6, 1.0, and 1.4 for  the \Bust signal; 
two \aetag ranges with boundaries at 0, 0.9, and 1.5 for the \Bdst signal, and 
two \aetag ranges with boundaries at 0, 0.8, and 1.4 for the \Bsst signal.

\begin{figure*}[th]
\centering
    \includegraphics[width=0.49\textwidth]{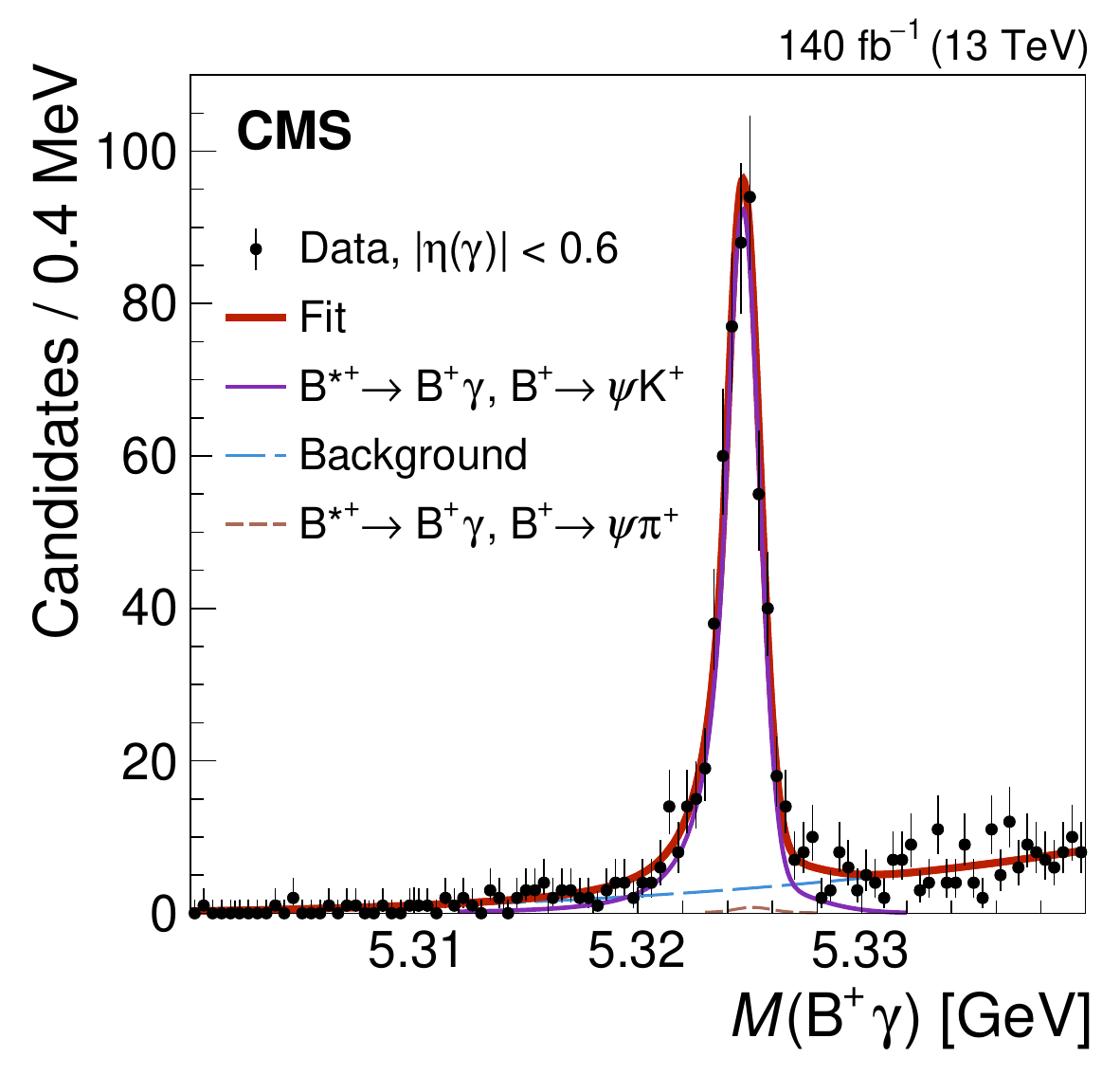}
    \includegraphics[width=0.49\textwidth]{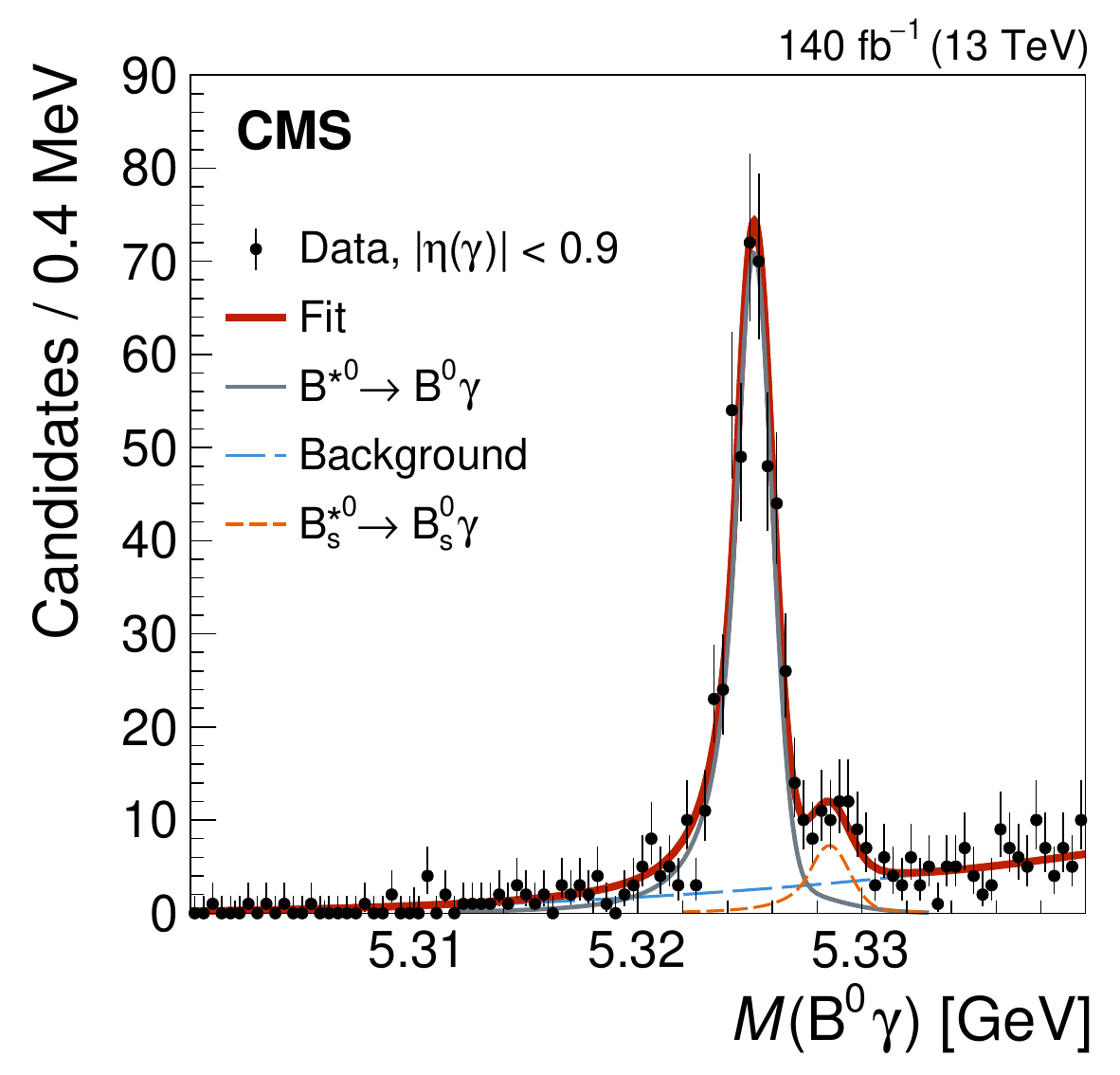}
    \includegraphics[width=0.49\textwidth]{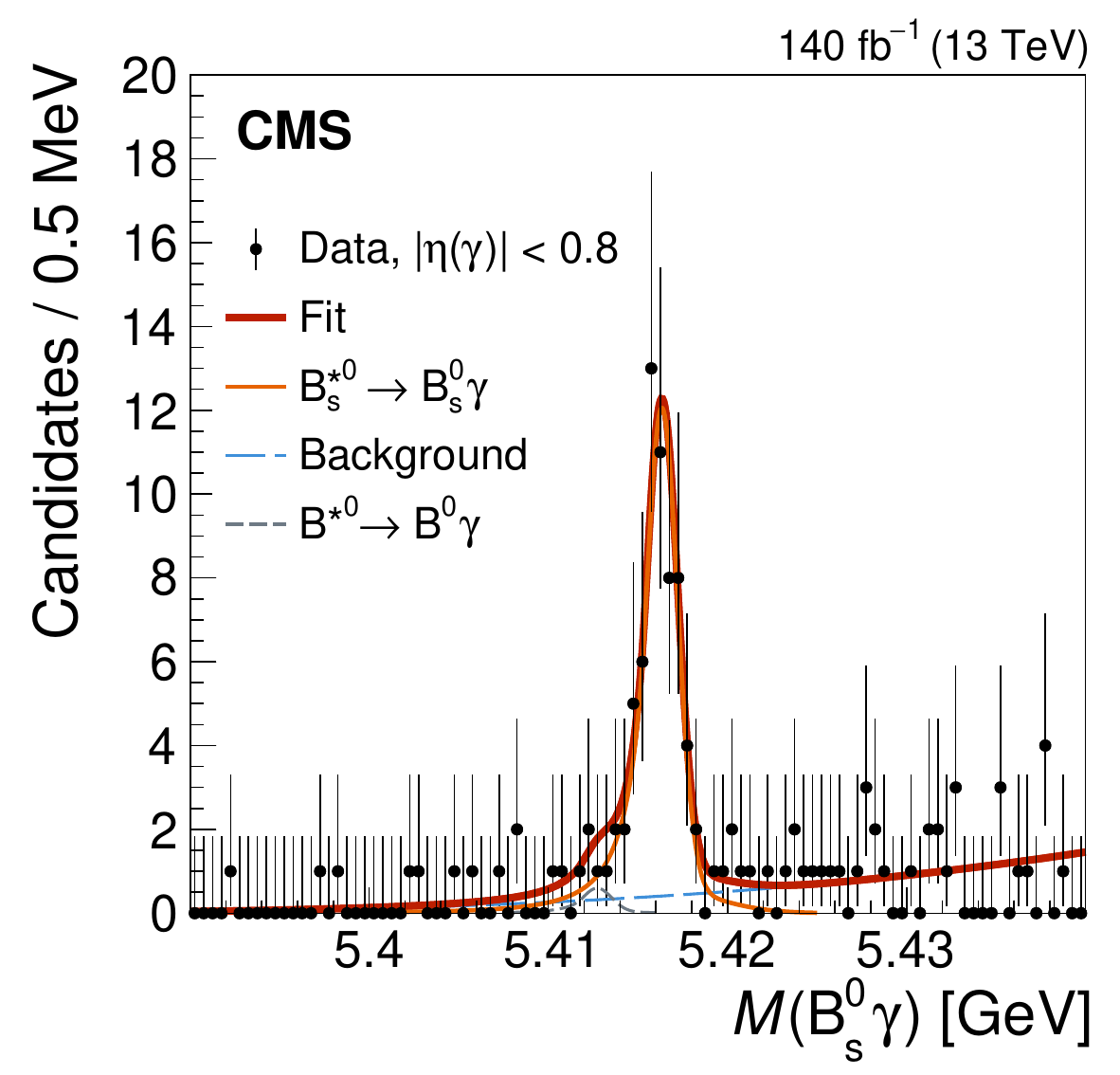}
    \caption{The measured distributions (filled circles) of the \Bug (upper left), 
    \Bdg (upper right), and \Bsg (lower) invariant mass in the 
    lowest \aetag range, which has the best invariant mass resolution.
    The simultaneous fit projections are shown with thick red lines, and the other
    components are described in the legends. The error bars represent statistical 
    uncertainties in data. \label{figfitmain}}
\end{figure*}

The measured invariant mass distributions of the selected \Bug, \Bdg, and
\Bsg candidates are shown in
Fig.~\ref{figfitmain} for the three categories with the best invariant mass resolution and in 
\suppMaterialB{} for the four remaining ones. 
A clear signal is seen in each of them.
A simultaneous extended unbinned maximum likelihood fit to the seven distributions is performed. 
Each signal is described by the sum of two Crystal Ball~(CB)~\cite{CB} functions, with common 
mean and tail parameters $n$ and $\alpha$. The three mean values, which are free parameters of the fit, 
correspond to the measured masses of the \Bust, \Bdst, and \Bsst mesons. The parameters $n$, $\alpha$, 
and the relative contribution of the two CB functions~($f$) are fixed to the ones obtained from simulation.
The widths of the CB functions ($\sigma_1$, $\sigma_2$) are multiplied by a common scale factor, which is a free parameter, 
to account for a possible data-simulation difference in the mass resolution.
For the \Bdst (\Bsst) signals, the yield in the high \aetag range is set to be 41\% (49\%) of the yield 
in the low \aetag range, as found in simulation. No such constraint is applied to the \Bust signal since it has been 
used to correct the \ptet distributions in simulation. 
The peaking contribution from the Cabibbo-suppressed \BuPsiPi decay in the \Bug mass spectrum is also modeled 
with the sum of two CB functions, with the shape parameters fixed to those found in simulation and the yield fixed 
relative to that for the \Bust signal, taking into account their relative branching fractions~\cite{PDG} and reconstruction efficiencies.
In the \Bdg mass distribution, the peaking contribution from $\BsstBsg\to\PsiKK\PGg$ decays (with one of 
the \PBzs daughter kaons reconstructed as a pion candidate from the \BdPsiKPi decay) 
is described by the sum of two CB functions with the shape parameters of the \Bsst signal 
and the mass shifted by $-87.78\MeV$ with respect to it, as found in the simulation.
The normalization of this component is left free in the fit. 
In the \Bsg mass distribution, the contribution from the \Bdst decays is described by the sum of two CB 
functions with the shape parameters of the \Bdst signal and the mass shifted by $+87.54\MeV$ with respect to it. 
Its normalization is fixed to 0.7\% of the \Bdst signal, according to the studies based on simulated event samples. 
Finally, the combinatorial background produced by random photons paired with a \PB meson 
is described in all distributions with a standard threshold function $(m-m_0)^{\beta}$, 
where $m_0$ is specific to each meson flavor, set to the corresponding ground-state meson mass, 
and $\beta$ is a free parameter, shared across the seven categories. 
Including all three meson flavors in a single fit allows for better control of cross contribution between 
\Bdst and \Bsst and more precise modeling of signal and background shapes that are expected to be the same.

The fit returns the mass differences $\dmBust\,\equiv\,\mBust-\mBu = 45.277\pm0.039\MeV$, 
$\dmBdst\,\equiv\,\mBdst-\mBd = 45.471\pm0.056\MeV$, and
$\dmBsst\,\equiv\,\mBsst-\mBs = 49.407\pm0.132\MeV$, where the uncertainties are statistical only.

The measurement uncertainties are dominated by the statistical component. 
The leading systematic uncertainties are discussed below. To evaluate the systematic uncertainties 
related to the choice of the signal model, several alternative functions are tested, with different 
$n$ or $\alpha$ in the CB functions, the sum of CB and Student's $t$ distributions, a CB 
function plus a Gaussian function, or a CB function plus a Novosibirsk~\cite{Belle:1999bhb} function.
In order to evaluate the uncertainties in the signal shape from the limited number of simulated events,
an alternative fit is performed, no longer fixing the shape parameters
but imposing that they are common to the (simultaneously fitted) measured and simulated distributions.
The nominal fit fixes the yield ratios between the \aetag categories for \Bdst and \Bsst signals 
to the values expected from simulation; alternatively, these constraints are removed.
Variations of the \BuJpsiPi contribution produce negligible changes in the results.
The peak positions and yields of the \Bsst to \Bdg and \Bdst to \Bsg cross-feed contributions are varied.
The alternative combinatorial background shape has seven independent $\beta$ parameters, one for each \aetag category.
For each of these sets of variations, the largest deviation in the measured quantities, with
respect to the results of the baseline fit, is taken as the corresponding systematic uncertainty.

A cross-check is performed by splitting the data into subsamples matching each of the three data-taking 
years (2016, 2017, and 2018) and measuring the corresponding \Bust mass values.
This cross-check also provides a good estimate of the uncertainty due to the detector alignment, 
since the pixel detector was replaced between 2016 and 2017. 
The results are found to be in good agreement with each other and the baseline result.

The statistical uncertainty from the PES correction is evaluated by making alternative PES 
corrections, with a momentum scale factor in each \aetag and \ptg range distributed as a Gaussian 
distribution, with mean equal to the baseline correction scale factor and RMS equal to its uncertainty. 
The simultaneous fit to the data is repeated each time, producing a set of \dmBst measurements. 
The RMS values of these distributions are taken as the systematic uncertainties in the measured mass differences.
The systematic uncertainty from the PES correction is evaluated by comparing the weighted average of the 
\piz mass values, obtained after the PES correction, in the \aetag and \ptg ranges relevant for the analysis 
with the world-average value of the \piz mass and scaling the difference by the ratio of \dmBst to the \piz mass.

All the systematic uncertainties described above are summarized in 
Table~\ref{tab:syst}, together with the total systematic uncertainty, 
calculated as the sum in quadrature of the individual components.
Systematic uncertainties in the various ratios or differences between the three reported mass differences
are evaluated in the same way and are given in \suppMaterialC.

\begin{table*}[th]
\centering
\topcaption{
Systematic uncertainties in the measured mass differences, in \keV. 
\label{tab:syst}}
\begin{scotch}{lccc}
    Source                          &  $\mBust-\mBu$    & $\mBdst-\mBd$     & $\mBsst-\mBs$ \\  
\hline
 Signal model                       &  \phantom{0}4     & \phantom{0}8      & 21            \\
 Signal shape parameters            &  15               & 15                & 18            \\
 Yield ratios of \aetag ranges      &  \phantom{0}1     & \phantom{0}2      & 10            \\
 Cross-feed $\Bsst\leftrightarrow\Bdst$                 
                                    &  $<$1\phantom{]}  & \phantom{0}1      & 10            \\
 Background shape                   &  \phantom{0}2     & $<$1\phantom{]}   & \phantom{0}7  \\
 Photon energy scale, statistical   &  12               & 14                & 16            \\
 Photon energy scale, systematic    &  18               & 18                & 19            \\[\cmsTabSkip]
 Total                              &  27               & 28                & 41                
\end{scotch}
\end{table*}

All the parameters measured in this analysis are presented in Table~\ref{tabres}. 
The mass differences (entries~1--3 in Table~\ref{tabres}) are consistent with, 
and an order of magnitude more precise than, the current world-average values~\cite{PDG}.
The measured mass difference $\mBsst-\mBs$ is in agreement 
with the Belle result~\cite{Belle:2008ezn}, and about two standard deviations away from the CLEO result~\cite{CLEO:2006yez}. 
The masses of the vector states (rows~4--6) are determined by adding 
the measured difference between the excited and ground states to the world-average values of the 
ground-state masses~\cite{PDG}. 

In addition, the measurements of the differences between the vector state masses (rows~7--10 in 
Table~\ref{tabres}), 
are reported, using the known mass differences between the \PB meson ground states~\cite{PDG}: 
$\mBd-\mBu=0.31\pm0.05\MeV$ and $\mBs-[\mBd+\mBu]/2=87.37\pm0.12\MeV$, 
assuming their uncertainties to be uncorrelated.
The result for $\mBdst-\mBust$ is consistent with the only previous measurement that used P-wave 
\PBzs mesons~\cite{CMS:2018wcx}, and is about 3 times more precise, while the other 
mass differences between vector states are measured for the first time. 
The differences and ratios between the mass differences are also measured for 
the first time (entries~11--18 in Table~\ref{tabres}), taking advantage of the 
cancellation of some uncertainties in these quantities.

\begin{table*}[thb]
\centering
\topcaption{Results of the measurement. The first uncertainties are statistical, 
the second are systematic, and the third, where present, are related to the masses 
or mass differences of the ground states taken from Ref.~\cite{PDG}. \label{tabres}}
\begin{scotch}{crl}
    & \multicolumn{1}{c}{Parameter}                 & \multicolumn{1}{c}{Value}                 \\  \hline
1   & $\dmBust\,\equiv\,\mBust-\mBu$                & $45.277\pm0.039\pm0.027\MeV$              \\
2   & $\dmBdst\,\equiv\,\mBdst-\mBd$                & $45.471\pm0.056\pm0.028\MeV$              \\
3   & $\dmBsst\,\equiv\,\mBsst-\mBs$                & $49.407\pm0.132\pm0.041\MeV$              \\[\cmsTabSkip]
4   & $\mBust$                                      & $5324.69\pm 0.04\pm 0.03\pm 0.07\MeV$     \\
5   & $\mBdst$                                      & $5325.19\pm 0.06\pm 0.03\pm 0.08\MeV$     \\
6   & $\mBsst$                                      & $5416.34\pm 0.13\pm 0.04\pm 0.10\MeV$     \\[\cmsTabSkip]
7   & $\mBdst-\mBust$                               & $\phantom{9}0.50\pm0.07\pm0.01\pm0.05\MeV$\\
8   & $\mBsst-\mBust$                               & $91.66\pm0.14\pm0.03\pm0.12\MeV$          \\
9   & $\mBsst-\mBdst$                               & $91.15\pm0.14\pm0.03\pm0.12\MeV$          \\
10  & $\mBsst-\frac{1}{2}\left[\mBdst+\mBust\right]$& $91.40\pm0.13\pm0.03\pm0.12\MeV$          \\[\cmsTabSkip]
11  & $\dmBdst-\dmBust$                             & $0.19\pm0.07\pm0.01\MeV$                  \\
12  & $\dmBsst-\dmBust$                             & $4.13\pm0.14\pm0.03\MeV$                  \\
13  & $\dmBsst-\dmBdst$                             & $3.94\pm0.14\pm0.03\MeV$                  \\
14  &$\dmBsst-\frac{1}{2}\left[\dmBdst+\dmBust\right]$&$4.03\pm0.13\pm0.03\MeV$                 \\[\cmsTabSkip]
15  & $\dmBdst/\dmBust$                             & $1.0043\pm0.0015\pm0.0002$                \\
16  & $\dmBsst/\dmBust$                             & $1.0912\pm0.0031\pm0.0007$                \\
17  & $\dmBsst/\dmBdst$                             & $1.0866\pm0.0031\pm0.0007$                \\
18  & $2\,\dmBsst/\left[\dmBust+\dmBdst\right]$     & $1.0889\pm0.0030\pm0.0007$                
\end{scotch}
\end{table*}

The measurements of \dmBsst, \dmBust, and \dmBdst agree with the predictions from lattice 
QCD~\cite{Bernardoni:2015nqa,Dowdall:2012ab,Gayer:2024akw}
for \dmBsst and $\dmBqst\,\equiv\,\mBqst-\mBq$ (where \Bq corresponds to either \PBp or \PBz since 
the predictions do not distinguish between the two), while the uncertainties in the predictions
are 1--2 orders of magnitude larger than those of the measurement. Phenomenological models 
provide more precise predictions including $\mBdst-\mBust=0.6\pm0.2$~\cite{Goity:2007fu} 
and $0.31\pm0.07\MeV$~\cite{Karliner:2019lau}, close to our measurement (entry~7 in Table~\ref{tabres}).
Reference~\cite{Goity:2007fu} also predicts hyperfine splittings $\dmBqst=45.69\pm0.02\MeV$ 
(above both values 1 and 2 in Table~\ref{tabres}) 
and $\dmBsst=46.73\pm0.08\MeV$ (below our measurement in line~3 of Table~\ref{tabres}). 
Experimental systematic uncertainties are reduced by taking the ratios of the mass differences, 
as shown in entries~16--18 in Table~\ref{tabres}. In addition, the theoretical uncertainties in the lattice 
QCD predictions from Ref.~\cite{Dowdall:2012ab} are also reduced, giving a prediction of $\dmBsst/\dmBqst=1.007\pm0.034$, 
which is about 2.5 standard deviations below our results and with an uncertainty that remains an order of magnitude larger.

In summary, three vector \PB meson states, \Bust, \Bdst, and \Bsst, have been fully reconstructed in 
exclusive final states for the first time by detecting low-energy photons through their conversion to \ee 
pairs in the beam pipe and detector material. The data sample of $\sqrt{s}=13\TeV$ proton-proton 
collisions was collected by the CMS experiment and corresponds to an integrated luminosity of 140\fbinv.
The measurements benefit from a new photon energy scale calibration method that uses \PizGG decays in 
which both photons convert to \ee pairs.
The masses of the three vector states are measured with respect to the corresponding ground states to be
$\mBust-\mBu = 45.277\pm0.039\pm0.027\MeV$, $\mBdst-\mBd = 45.471\pm0.056\pm0.028\MeV$, and
$\mBsst-\mBs = 49.407\pm0.132\pm0.041\MeV$, where the first uncertainties are statistical and the second are 
systematic. 
A number of difference and ratio measurements between the reported masses and mass 
differences are also provided, where, due to cancellations in the computations of these quantities, 
both experimental and theoretical uncertainties are lower. 
These results are either the first measurement or an order of magnitude more precise than previous measurements.
The new measurements are more precise than the available theoretical predictions and provide an important 
input to our understanding of heavy-quark systems.

\begin{acknowledgments}
We congratulate our colleagues in the CERN accelerator departments for the excellent performance of the LHC and thank the technical and administrative staffs at CERN and at other CMS institutes for their contributions to the success of the CMS effort. In addition, we gratefully acknowledge the computing centers and personnel of the Worldwide LHC Computing Grid and other centers for delivering so effectively the computing infrastructure essential to our analyses. Finally, we acknowledge the enduring support for the construction and operation of the LHC, the CMS detector, and the supporting computing infrastructure provided by the following funding agencies: SC (Armenia), BMBWF and FWF (Austria); FNRS and FWO (Belgium); CNPq, CAPES, FAPERJ, FAPERGS, and FAPESP (Brazil); MES and BNSF (Bulgaria); CERN; CAS, MoST, and NSFC (China); MINCIENCIAS (Colombia); MSES and CSF (Croatia); RIF (Cyprus); SENESCYT (Ecuador); ERC PRG, TARISTU24-TK10 and MoER TK202 (Estonia); Academy of Finland, MEC, and HIP (Finland); CEA and CNRS/IN2P3 (France); SRNSF (Georgia); BMBF, DFG, and HGF (Germany); GSRI (Greece); NKFIH (Hungary); DAE and DST (India); IPM (Iran); SFI (Ireland); INFN (Italy); MSIT and NRF (Republic of Korea); MES (Latvia); LMTLT (Lithuania); MOE and UM (Malaysia); BUAP, CINVESTAV, CONACYT, LNS, SEP, and UASLP-FAI (Mexico); MOS (Montenegro); MBIE (New Zealand); PAEC (Pakistan); MES and NSC (Poland); FCT (Portugal);  MESTD (Serbia); MICIU/AEI and PCTI (Spain); MOSTR (Sri Lanka); Swiss Funding Agencies (Switzerland); MST (Taipei); MHESI and NSTDA (Thailand); TUBITAK and TENMAK (T\"{u}rkiye); NASU (Ukraine); STFC (United Kingdom); DOE and NSF (USA).
\end{acknowledgments}

\bibliography{auto_generated}

\ifthenelse{\boolean{cms@external}}{}{
\clearpage
\numberwithin{table}{section}
\numberwithin{figure}{section}
\appendix
\section{Example distributions of the \texorpdfstring{$\PizGG$}{pi0 to gamma gamma} invariant mass\label{app:suppMat}}
The distributions of the \GG invariant mass from two converted photons in 2018 data, showing 
the \PizGG peak in the kinematic range $0.8<\ptg<0.9\gev$, $0.8<\aetag<0.9$,
are presented in Fig.~\ref{figpiz} before (left) and after (right) the PES correction. 
The distributions are fit with a sum of two Crystal Ball functions with a common mean for the signal and 
a polynomial multiplied by a power law function for the background. 

\begin{figure*}[ht]
\centering
    \includegraphics[width=0.48\textwidth]{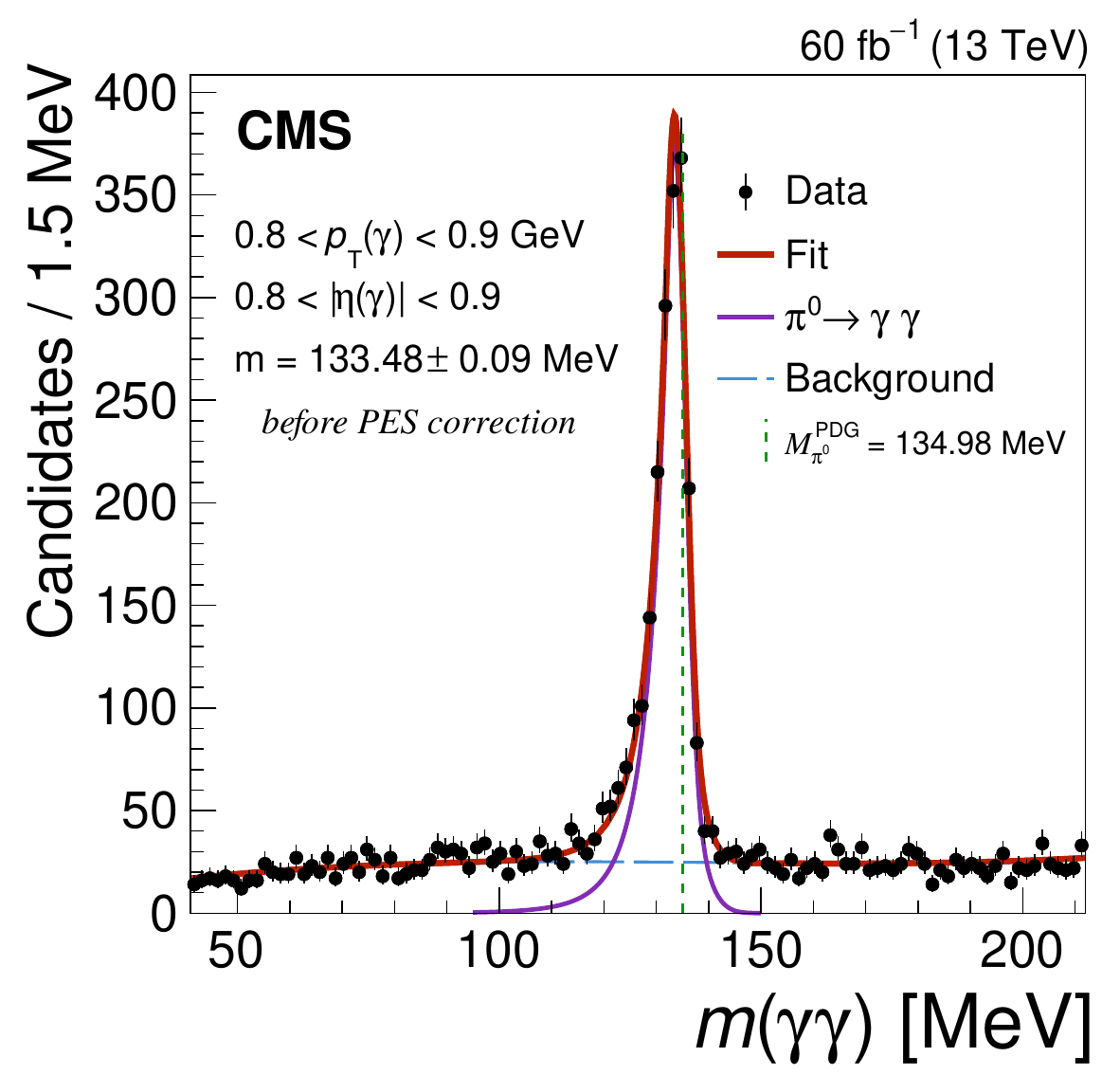}
    \includegraphics[width=0.48\textwidth]{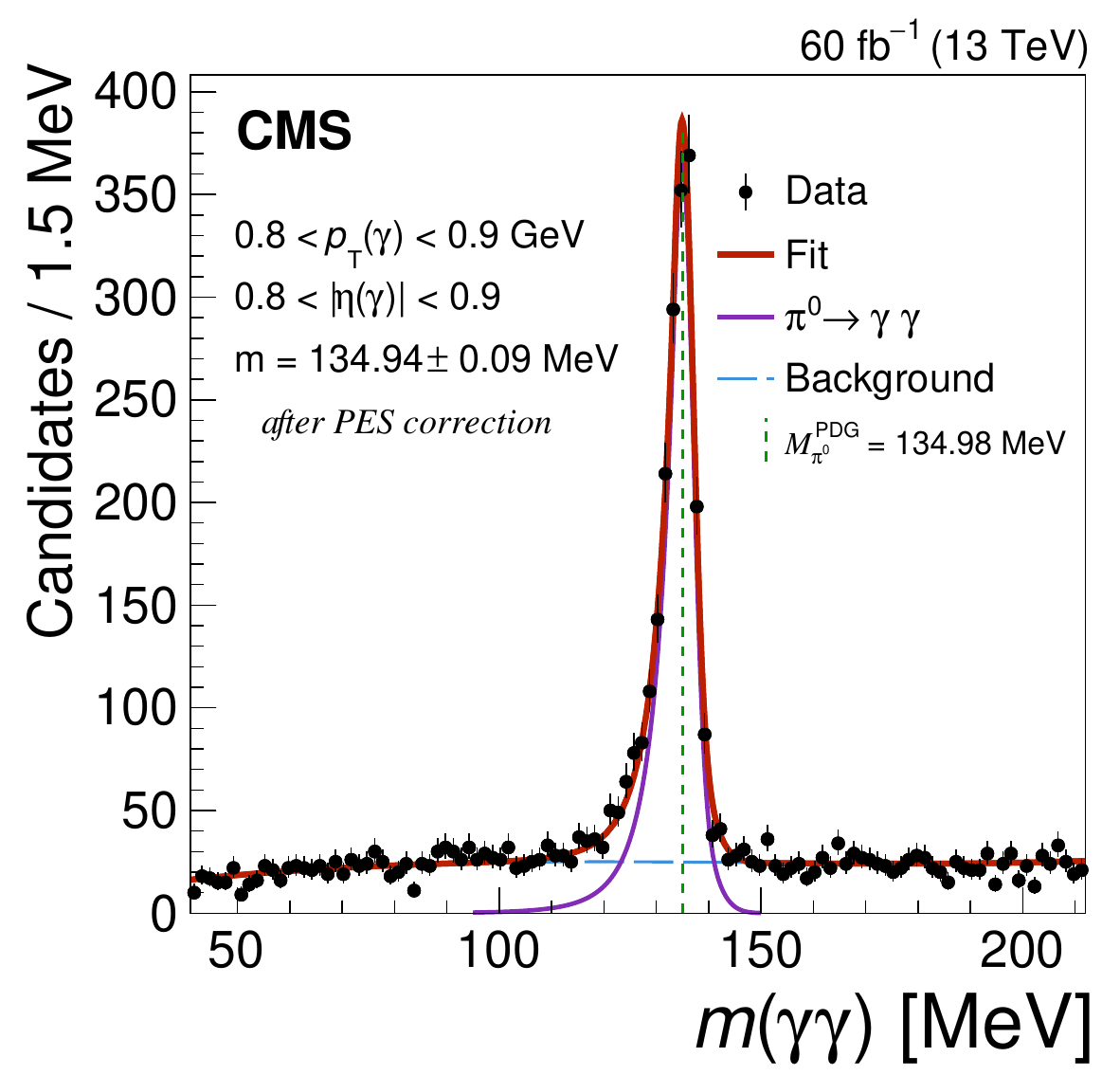}
    \caption{The measured distributions of the diphoton invariant mass near the \piz mass with 
    the fit results overlaid, before (left) and after (right) the PES correction, in the kinematic range 
    $0.8<\ptg<0.9\gev$, $0.8<\aetag<0.9$. The green vertical line shows the world-average value of the \piz
    mass of 134.98\MeV~\cite{PDG}. \label{figpiz}}
\end{figure*}

\section{Fit projections in extra pseudorapidity ranges\label{app:suppMatB}}
The measured invariant mass distributions of the \Bug, \Bdg, and \Bsg candidates 
in the four additional categories of the simultaneous fit are shown in
Fig.~\ref{figfitapp}.

\begin{figure*}[ht]
\centering
    \includegraphics[width=0.48\textwidth]{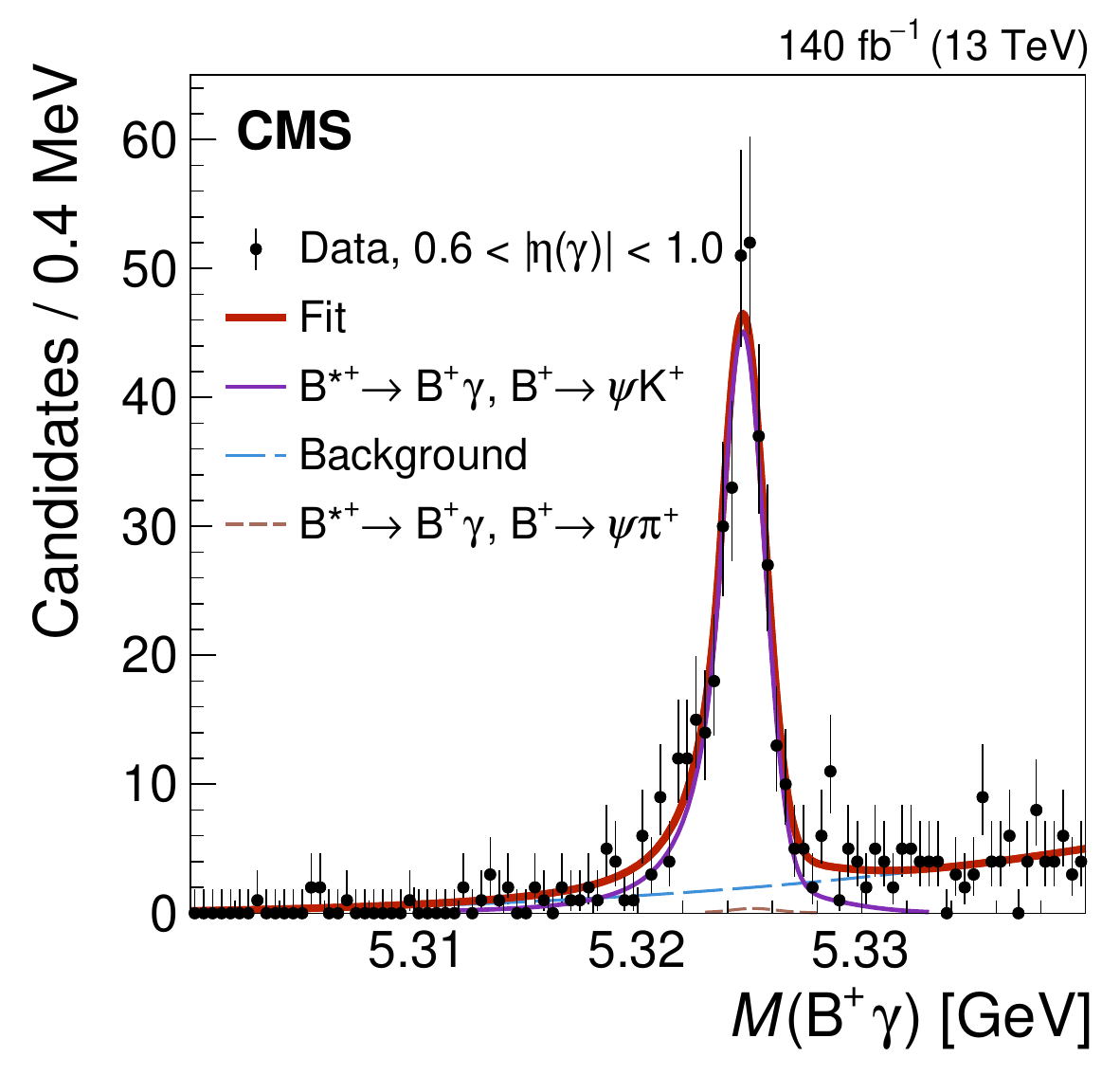}
    \includegraphics[width=0.48\textwidth]{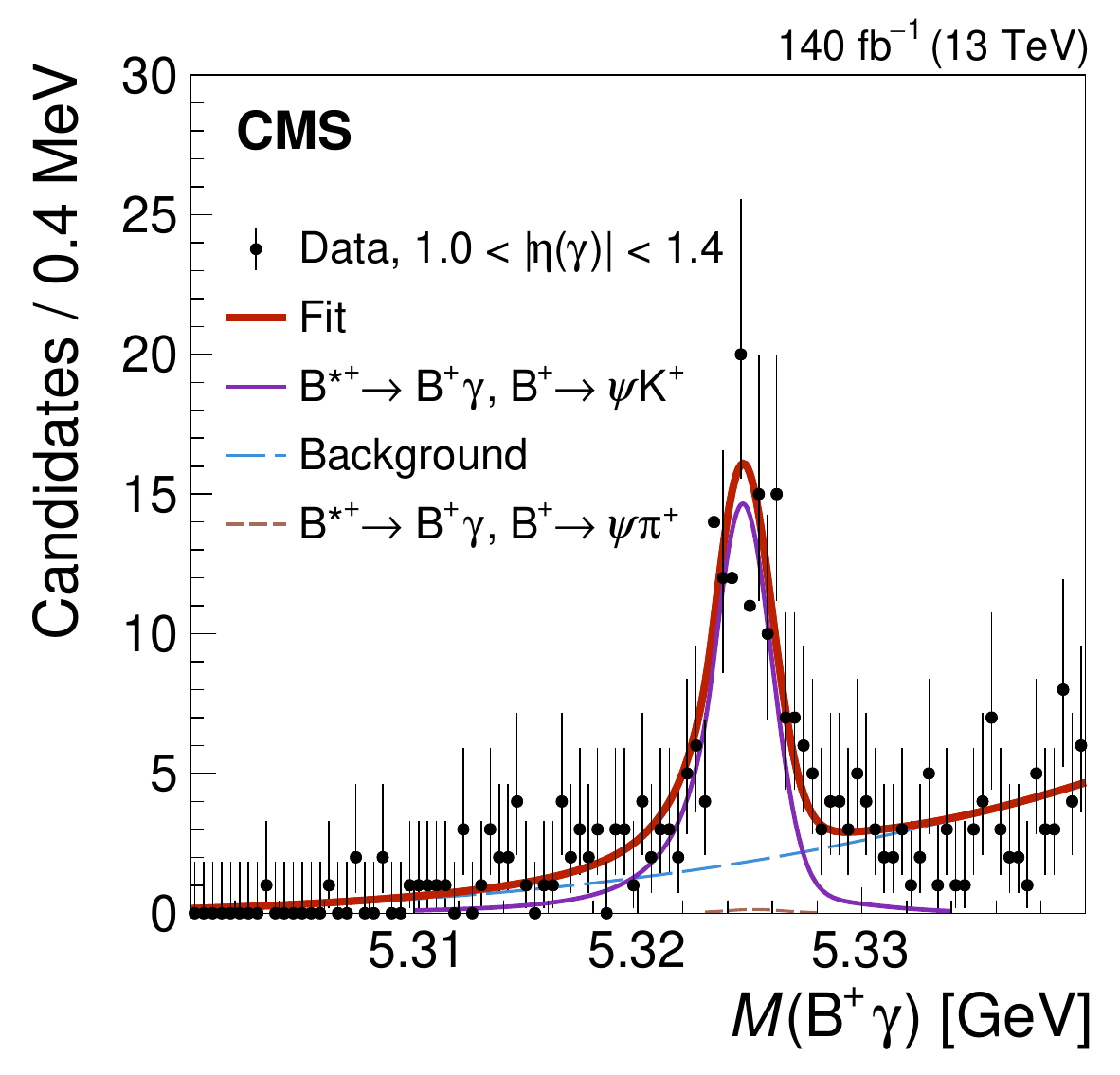}
    \includegraphics[width=0.48\textwidth]{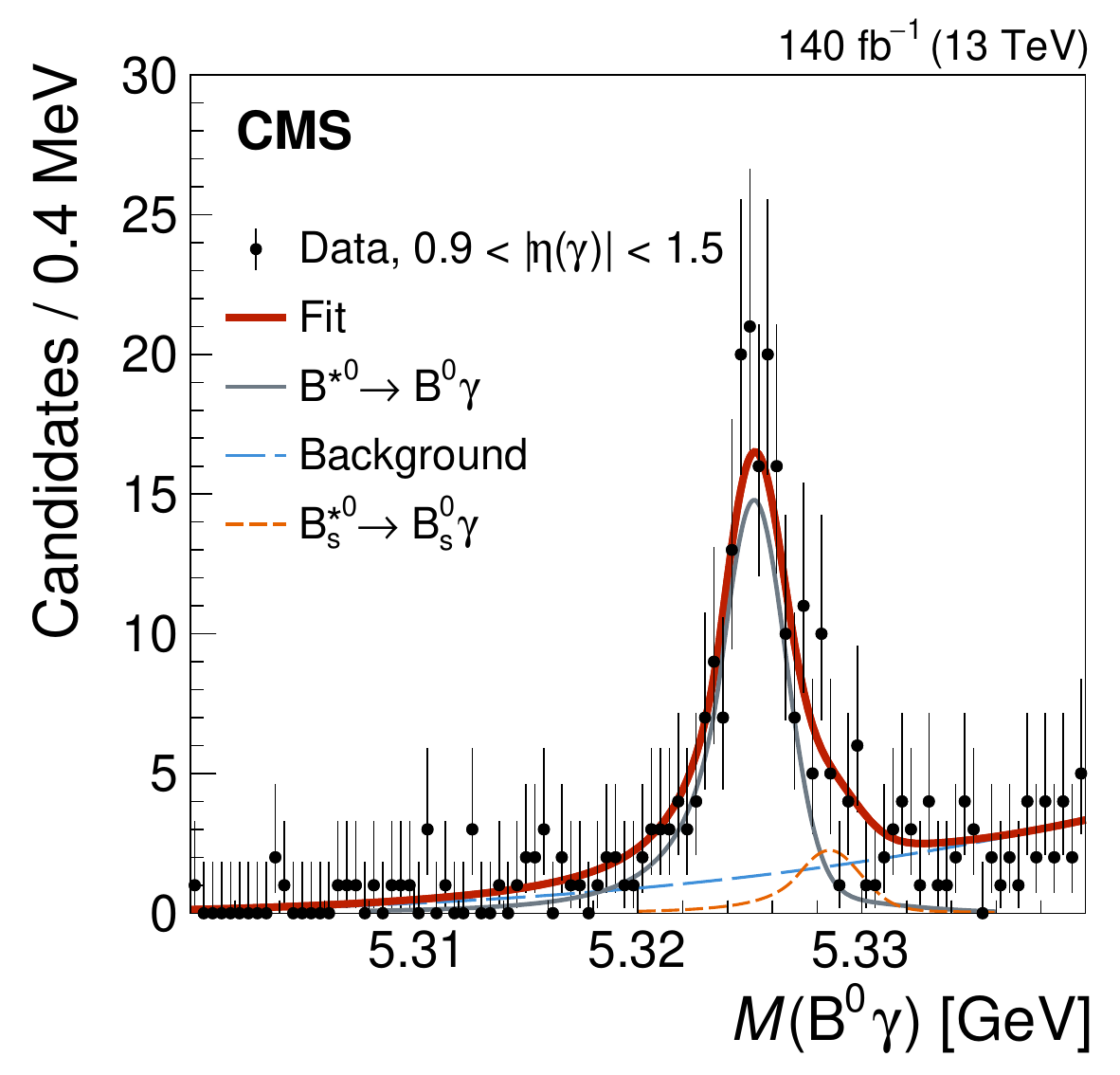}
    \includegraphics[width=0.48\textwidth]{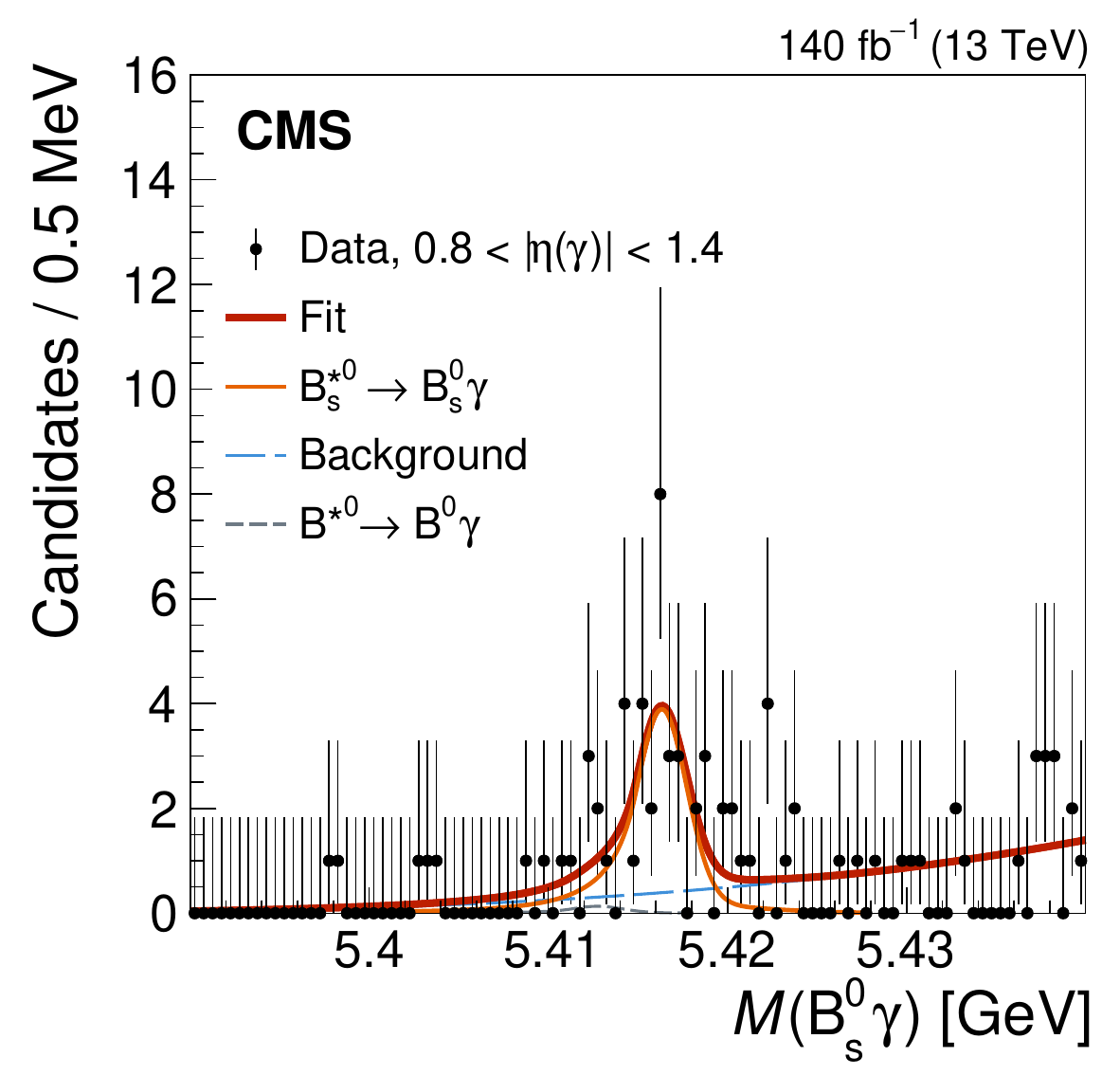}
    \caption{The measured distributions (filled circles) of the \Bug (upper), 
    \Bdg (lower left), and \Bsg (lower right) invariant mass corresponding 
    to the most forward one or two \aetag regions,
    together with the simultaneous fit projections (thick red lines).
    The error bars represent statistical uncertainties in data.\label{figfitapp}}
\end{figure*}

\section{Systematic uncertainties in the differences and ratios of the mass differences\label{app:suppMatC}}

The systematic uncertainties for the differences and ratios of the mass 
differences (entries~11--18 in the tabulated list of results) are evaluated in the 
same manner as for the \dmBst uncertainties, with the variations to the baseline 
model applied simultaneously to all channels to account for correlations. 
The resulting uncertainties are shown in Tables~\ref{tab:systextrad} and~\ref{tab:systextrar}.
The experimental statistical and systematic uncertainties for lines~7--10 of the 
tabulated list of results are identical to those for lines~11--14 (Table~\ref{tab:systextrad}).
As for the other measurements, the statistical uncertainties dominate. 
Variations of photon energy scale produce correlated variations in the \dmBsst, \dmBust, 
and \dmBdst measurements. Consequently, the corresponding systematic uncertainties are partially (fully)
canceled in the measured differences (ratios) of the mass differences.

\begin{table*}[ht]
\centering
\topcaption{
Values, statistical, and systematic uncertainties, in \keV, of the measured differences of the three mass differences.
\label{tab:systextrad}}
\cmsTabla{\begin{scotch}{lcccc}
    Source                              & $\dmBdst-\dmBust$ & $\dmBsst-\dmBust$ & $\dmBsst-\dmBdst$ & $\dmBsst-\frac{\dmBdst+\dmBust}{2}$ \\[\cmsTabSkip]
\hline
    Measured value                      & 194               & 4130              & 3936              & 4033              \\
    Statistical uncertainty             & \phantom{0}68     & \phantom{0}138    & \phantom{0}139    & \phantom{0}134    \\
\hline 
 Signal model                           & \phantom{00}4     & \phantom{00}23    & \phantom{00}23    & \phantom{00}23    \\
 Signal shape parameters                & \phantom{00}1     & \phantom{000}3    & \phantom{000}3    & \phantom{000}3    \\
 Yield ratios of \aetag ranges          & \phantom{00}3     & \phantom{00}11    & \phantom{000}7    & \phantom{000}9    \\
 Cross-feed $\Bsst\leftrightarrow\Bdst$ & \phantom{00}1     & \phantom{00}10    & \phantom{00}12    & \phantom{00}11    \\
 Background shape                       & \phantom{00}3     & \phantom{000}9    & \phantom{000}6    & \phantom{000}8    \\
 Photon energy scale, statistical       & \phantom{00}4     & \phantom{00}11    & \phantom{00}11    & \phantom{00}11    \\
 Photon energy scale, systematic        & \phantom{0}$<$1\phantom{]} &\phantom{000}2 & \phantom{000}2 & \phantom{000}2  \\[\cmsTabSkip]
 \hline 
 Total systematic uncertainty           & \phantom{00}7     & \phantom{00}31    & \phantom{00}30    & \phantom{00}30
\end{scotch}}
\end{table*}

\begin{table*}[ht]
\centering
\topcaption{
Values, statistical, and systematic uncertainties of the measured ratios of the three mass differences.
\label{tab:systextrar}}
\cmsTable{\begin{scotch}{lcccc}
    Source                              & $\dmBdst/\dmBust$ & $\dmBsst/\dmBust$ & $\dmBsst/\dmBdst$ & $\frac{2\,\dmBsst}{\dmBust+\dmBdst}$ \\[\cmsTabSkip]
\hline
    Measured value                      & 1.00428           & 1.09122           & 1.08656           & 1.08888                           \\
    Statistical uncertainty             & 0.00151           & 0.00306           & 0.00309           & 0.00297                           \\
\hline 
 Signal model                           & 0.00009           & 0.00050           & 0.00052           & 0.00052                           \\ 
 Signal shape parameters                & 0.00002           & 0.00003           & 0.00004           & 0.00003                           \\ 
 Yield ratios of \aetag ranges          & 0.00008           & 0.00023           & 0.00016           & 0.00020                           \\
 Cross-feed $\Bsst\leftrightarrow\Bdst$ & 0.00003           & 0.00023           & 0.00025           & 0.00015                           \\
 Background shape                       & 0.00005           & 0.00020           & 0.00014           & 0.00017                           \\
 Photon energy scale (statistical)      & 0.00009           & 0.00025           & 0.00024           & 0.00024                           \\[\cmsTabSkip]
 \hline 
 Total systematic uncertainty           & 0.00016           & 0.00067           & 0.00066           & 0.00065     
\end{scotch}}
\end{table*}

}
\cleardoublepage \section{The CMS Collaboration \label{app:collab}}\begin{sloppypar}\hyphenpenalty=5000\widowpenalty=500\clubpenalty=5000\cmsinstitute{Yerevan Physics Institute, Yerevan, Armenia}
{\tolerance=6000
A.~Hayrapetyan, V.~Makarenko\cmsorcid{0000-0002-8406-8605}, A.~Tumasyan\cmsAuthorMark{1}\cmsorcid{0009-0000-0684-6742}
\par}
\cmsinstitute{Institut f\"{u}r Hochenergiephysik, Vienna, Austria}
{\tolerance=6000
W.~Adam\cmsorcid{0000-0001-9099-4341}, J.W.~Andrejkovic, L.~Benato\cmsorcid{0000-0001-5135-7489}, T.~Bergauer\cmsorcid{0000-0002-5786-0293}, M.~Dragicevic\cmsorcid{0000-0003-1967-6783}, C.~Giordano, P.S.~Hussain\cmsorcid{0000-0002-4825-5278}, M.~Jeitler\cmsAuthorMark{2}\cmsorcid{0000-0002-5141-9560}, N.~Krammer\cmsorcid{0000-0002-0548-0985}, A.~Li\cmsorcid{0000-0002-4547-116X}, D.~Liko\cmsorcid{0000-0002-3380-473X}, M.~Matthewman, I.~Mikulec\cmsorcid{0000-0003-0385-2746}, J.~Schieck\cmsAuthorMark{2}\cmsorcid{0000-0002-1058-8093}, R.~Sch\"{o}fbeck\cmsAuthorMark{2}\cmsorcid{0000-0002-2332-8784}, D.~Schwarz\cmsorcid{0000-0002-3821-7331}, M.~Shooshtari, M.~Sonawane\cmsorcid{0000-0003-0510-7010}, W.~Waltenberger\cmsorcid{0000-0002-6215-7228}, C.-E.~Wulz\cmsAuthorMark{2}\cmsorcid{0000-0001-9226-5812}
% ...
\par}
\vskip\cmsinstskip
\dag:~Deceased\\
$^{1}$Also at Yerevan State University, Yerevan, Armenia\\
$^{2}$Also at TU Wien, Vienna, Austria\\
% ...
\end{sloppypar}
%%% END EDITABLE REGION %%%
% skeleton_end
\end{document}